\documentclass[a4paper,12pt]{article}
\usepackage{amsmath,amsfonts,amssymb,cite}
\usepackage[dvips]{graphicx}

\begin{document}

\begin{center}
{\LARGE\bf Lagrangian alternative to QCD string
%A Lagrangian description of the radial Regge spectrum
}
\end{center}

\begin{center}
{\large S. S. Afonin and A. D. Katanaeva}
\end{center}

\begin{center}
{\it V. A. Fock Department of Theoretical Physics, Saint
Petersburg State University, 1 ul. Ulyanovskaya, 198504, Russia
\\ \rm{E-mail: afonin@hep.phys.spbu.ru}}
\end{center}

\begin{abstract}
The spectrum of radially excited hadrons provides much information
about the confinement forces in QCD. The confinement is realized
most naturally in terms of the QCD string whose quantization gives
rise to the radially excited modes. We propose an alternative
framework for the description of the excited spectrum. Namely we
put forward some effective field models in which the hadrons
acquire masses due to interaction with a scalar field modeling the
non-perturbative gluon vacuum. The effective potential for this
field is periodic with an infinite number of non-equivalent vacua.
The radially excited hadrons emerge as elementary excitations over
different vacua. We construct explicit examples for such effective
theories in the meson sector. An interesting byproduct of the
considered models is the existence of a classical field
configuration in each vacuum. Depending on the model, it can
represent a domain wall, Nielsen-Olesen vortex or
't~Hooft--Polyakov monopole. In the pure scalar sector, it is
shown that the first quantum correction leads to splitting of the
radially excited modes into two nearly degenerate states. Such a
phenomenon has some phenomenological support. The presented
approach may be also viewed as an alternative framework to the bottom-up
AdS/QCD models where the radial excitations appear on the
classical level as well.
\end{abstract}

%\newpage

\section{Introduction}

A large number of observed hadron resonances can be interpreted as
the radial excitations of some lighter hadrons due to identical
quantum numbers. The form of spectrum of the radial excitations is
determined by the confinement forces in Quantum Chromodynamics
(QCD). Thus the study of radial excitations seems to be
indispensable to the understanding of the confinement dynamics.

The quark confinement and the radial excitations arise naturally in
the hadron string models where the string serves as an approximation
for the gluon flux-tube with nearly massless quarks at the
ends~\cite{nussinov,string}. This relativistic approach is well
accommodated for the description of light mesons representing
ultrarelativistic systems. The hadron strings usually imply that the
limit of large number of colors $N_c$ in QCD~\cite{hoof} is taken.
The meson excitations become narrow, and hence well defined objects
for the theoretical study, with the masses being approximately
independent of $N_c$. In addition, the radial spectrum must contain
an infinite number of states. The semiclassical quantization of a
thin flux-tube typically gives a linearly rising spectrum for masses
squared~\cite{string,shifman},
\begin{equation}
\label{1}
M_k^2=a(k+b), \qquad k=0,1,2,\dots,
\end{equation}
where the slope $a$ is proportional to the string tension. Since the
latter is governed by the gluodynamics, the slope $a$ is expected to
be unaffected by the quantum numbers of meson resonances. The known
spectrum of light mesons seems to agree with the
behavior~\eqref{1}~\cite{phen,klempt}.

The concept that QCD in the large-$N_c$ limit may be formulated as
some string theory has been evolving for some decades. The usual
models based on the QCD string possess a nice basic feature: The
linear Regge recurrence, $m_J^2\sim J$ ($J$ denotes the hadron
spin), emerges already on the classical level of rotating string.
The radial modes appear after quantization. Unfortunately, a
consistent theory of quantized QCD string has not been built in
spite of extensive efforts. It is therefore interesting to look for
alternative dynamical approaches leading to the relation~\eqref{1}
in some natural way, at least approximately. The term "natural" we
understand in a specific way: It would be interesting to construct a
model where the relation~\eqref{1} holds on the classical level,
like the relation $m_J^2\sim J$ in the rotating string. In this
case, one can hope to escape the problems related to the rigorous
quantization. In some sense, an example of such a model is given by
the five-dimensional soft-wall holographic model~\cite{son2}. We
wish to construct an example staying within the four-dimensional
framework.

In fact, a careful analysis of existing data shows that a stronger
statement than the recurrence~\eqref{1} can be made: The light
non-strange mesons cluster near certain equidistant values of
energy squared with repeating structure of the
clusters~\cite{klempt,Cluster,clust_rev,cluster,glozman}, see,
{\it e.g.}, figure in Appendix A. The matter looks as if the
underlying dynamics recurred periodically when one moves to higher
energies. This observation suggests that as a starting point for
the model building we may take an assumption that the gluon vacuum
has the periodic structure in a dynamical sense\footnote{Here we
mean the periodicity in energy scales. This should not be confused
with the periodic structure of topological vacua in the pure
Yang-Mills theories where the tunneling between vacua occurs due
to the instantons. We do not know whether there is any relation
between the topological vacua ({\it e.g.}, because of the presence
of quarks) and the dynamical vacua under consideration.}.

In the present paper, we put forward a new dynamical approach to
the description of the radial hadron excitations. The approach is
based on a realization of the aforementioned assumption within the
framework of an effective field theory. We construct models in
which the mesons acquire masses via a Higgs-like mechanism due to
interaction with a scalar field $\varphi$. The central idea is
that the effective potential for $\varphi$ is periodic and its
vacua are non-equivalent. This leads to the appearance of infinite
set of vacuum expectation values $\langle\varphi\rangle_k$,
$k=0,1,2,\dots$. Correspondingly, the hadron fields coupled to
$\varphi$ can have different mass terms depending  on the choice
of $\langle\varphi\rangle_k$, {\it i.e.} depending on the scale we
probe the theory. As a consequence, each hadron has an infinite
discrete set of masses $M_k$, the states with $k>0$ are identified
with the radial excitations of the ground state having $k=0$.

Needless to say that this approach proposes a new viewpoint on the
nature of the radially excited hadrons. Perhaps the following
mechanical analogy clarifies the essence of our idea. Consider the
mathematical pendulum with a friction. Depending on the energy of
push, it will oscillate near its equilibrium position after $k$
turns. The energy of hadron creation corresponds to the energy of
push in this rough analogy and the "winding number" $k$ does to
the radial number of the created hadron. An interesting byproduct
of the considered models is the existence of non-trivial classical
solutions which will be analyzed in detail.

The paper is organized as follows. In Sect.~2, we construct an
effective model of single scalar field which possesses an infinite
number of non-degenerate vacua. This model is used in Sect.~3 to
describe the radial spectrum of the vector mesons. The emerging
vortex-line solutions are discussed in detail. The quantum
correction to the scalar model of Sect.~2 is analyzed in Sect.~4.
The Sect.~5 contains discussions concerning the phenomenology of
the proposed model and possible connections of our approach with
effective models for QCD. Our conclusions and some prospects for
future work are presented in Sect.~6.

\section{Scalar theory with non-degenerate vacua}

Consider the following effective field theory for a real scalar
field in four dimensions,
\begin{equation}
\label{2}
\mathcal{L}=\frac{1}{2}\partial_ \mu\varphi\partial^\mu\varphi+
\frac{\mu^4}{2\lambda}\left[\cos\left(\frac{\lambda}{\mu^2}\varphi^2-2\pi b\right)-\cos(2\pi b)\right],
\end{equation}
where $\mu$ has the dimension of mass and the parameters $\lambda$
and $b$ are dimensionless. The periodicity of the interaction term
in~\eqref{2} imposes a restriction on the value of parameter $b$:
the interval $0\leq b< 1$ covers all nontrivial cases. For instance,
the case $b=1$ is equivalent to the $b=0$ one, {\it etc}. The
parameter $\lambda$ plays the role of coupling constant. In the weak
coupling regime, $|\lambda|\ll1$, the Lagrangian~\eqref{2} reduces
to the scalar theory $\lambda\varphi^4$,
\begin{equation}
\label{3}
\mathcal{L}_{|\lambda|\ll1}=\frac{1}{2}\partial_\mu\varphi\partial^\mu\varphi+\frac12\mu^2\sin(2\pi
b)\varphi^2- \frac{\lambda}{4}\cos(2\pi
b)\varphi^4+\mathcal{O}(\lambda^2\varphi^6).
\end{equation}
As is seen from~\eqref{3}, the parameter $b$ induces the mass term
except the cases $b=0$ and $b=\frac12$. We could introduce the mass
term $\frac12m^2\varphi^2$ from the very beginning and set $b=0$.
This theory would be completely equivalent. We find the
variant~\eqref{2} more convenient for our purposes because the
parameter $b$ will have then a direct meaning of intercept in the
relation~\eqref{1}. The expansion~\eqref{3} shows also that the
energy functional is bounded from below in the weak coupling regime
if $\lambda\cos(2\pi b)>0$. This results in the following
interrelation between the sign of the coupling $\lambda$ and the
value of $b$: $0\leq b<\frac14$ or $\frac34\leq b<1$ for $\lambda>0$
and $\frac14\leq b<\frac34$ for $\lambda<0$.

At $b=0$ the Lagrangian~\eqref{2} resembles the sine-Gordon
model~\cite{ra}. The model~\eqref{2}, however, is quite different
since the nonlinear power of $\varphi$ in the cosine destroys the translational
invariance (modulo a general factor) $\varphi\rightarrow\varphi+2\pi
k$, $k=0,\pm1,\pm2,\dots$. The vacua are therefore non-equivalent,
although they deliver equal minima to the action.

The potential in~\eqref{2} is minimized on the constant field configurations
\begin{equation}
\label{4}
\langle\varphi\rangle_k=\sqrt{\frac{2\pi\mu^2}{\lambda}(k+b)}, \qquad k=0,1,2,\dots.
\end{equation}
Consider small perturbations around the vacua~\eqref{4},
\begin{equation}
\label{5}
\varphi=\langle\varphi\rangle_k+\sigma(x).
\end{equation}
The quadratic part of the Lagrangian for $\sigma$ reads
\begin{equation}
\label{6}
\mathcal{L}_{\sigma}^{(2)}=\frac12\partial_\mu\sigma\partial^\mu\sigma-\lambda\langle\varphi\rangle_k^2\sigma^2.
\end{equation}
The spectrum of excitations follows from~\eqref{6} and~\eqref{4},
\begin{equation}
\label{7}
M^2_{\sigma,k}=2\lambda\langle\varphi\rangle_k^2=4\pi\mu^2(k+b),
\end{equation}
which has the Regge-like form~\eqref{1}. Thus the considered model
has an infinite number of unconnected vacua with different
excitation energies given by the relation~\eqref{7}.

It is worth mentioning that a simple semiclassical quantization of
the gluon string with massless quarks confined by the linear
potential leads to the relation~\cite{shifman}: $M^2_k=4\pi\tau
k+\text{const}$, where $\tau$ denotes the string tension. The mass
parameter $\mu^2$ has thus an analogy with the string tension.

The seeming similarity of the Lagrangian~\eqref{2} with the
sine-Gordon model poses an interesting question whether there
exist some soliton-like solutions in the two-dimensional case. In
fact, the most close analogue to the sine-Gordon model is provided
by the following modification of the Lagrangian~\eqref{2},
\begin{equation}
\label{8}
\tilde{\mathcal{L}}=\frac{1}{2}\partial_
\mu\varphi\partial^\mu\varphi+\frac{\mu^6}{\lambda^2\varphi^2}\left[
\cos\left(\frac{\lambda}{\mu^2}\varphi^2\right)-1\right].
\end{equation}
The minima of effective potentials in~\eqref{8} and in~\eqref{2} at
$b=0$ coincide. As $\lambda\rightarrow0$ we obtain the theory
\begin{equation}
\label{9}
\tilde{\mathcal{L}}=\frac{1}{2}\partial_
\mu\varphi\partial^\mu\varphi-
\frac{1}{2}\mu^2\varphi^2+\frac{1}{4!}\frac{\lambda^2}{\mu^2}\varphi^6+\dots,
\end{equation}
where the mass term arises from the cosine as in the sine-Gordon
model~\cite{ra}. The equation of motion for the field $\varphi$ in
the Lagrangian~\eqref{8} has a soliton-like solution which in the
static case takes the form
\begin{equation}
\label{10}
\varphi=\pm2\sqrt{\frac{\mu^2}{\lambda}\arctan{e^{2\mu(x-x_0)}}}.
\end{equation}
The expression in~\eqref{10} represents nothing but the square
root of the static one-soliton solution in the sine-Gordon
model~\cite{ra}. The solution~\eqref{10}, however, is not a
genuine soliton\footnote{By definition, the genuine solitons do
not change their shapes after interactions with another
solitons~\cite{ra}. If a non-linear differential equation has a
"genuine" one-soliton solution then it necessarily possesses the
two-soliton, three-soliton {\it etc.} solutions. The $(n+1)$-th
soliton solution can be obtained from the $n$-th one by means of
the so-called B\"{a}cklund transformations (a kind of the
principle of the non-linear superposition). The absence of these
transformations entails the absence of "genuine" solitons.}. This
conclusion follows from the absence of the B\"{a}cklund
transformations for the system~\eqref{8}. In the relativistic
systems described by the Lagrangians
$\mathcal{L}=\frac12(\partial_ \mu\varphi)^2-V(\varphi)$, the
necessary condition for the existence of the B\"{a}cklund
transformations is~\cite{backlund}: $V'''+\alpha^2V'=0$, where
$\alpha$ is a parameter and the prime denotes derivative. This
condition is not fulfilled in the model~\eqref{8}. The
solution~\eqref{10} describes a usual kink, which connects two
neighbouring vacua, and in four dimensions it becomes a domain
wall. The same situation takes place in the model~\eqref{2}, where
the kinks exist but cannot be found analytically.

\section{Abelian Higgs model with periodic potential}

Consider now a $U(1)$ gauge model of the Nielsen-Olesen
type~\cite{nielsen} in which the Higgs potential is taken from the
scalar model~\eqref{2}. The Lagrangian reads
\begin{equation}
\label{11}
\mathcal{L}=D_\mu\varphi(D^\mu\varphi)^*+
\frac{\mu^4}{2\lambda}\left[\cos\left(\frac{2\lambda}{\mu^2}\varphi\varphi^*-2\pi b\right)
-\cos(2\pi b)\right]-\frac14F_{\mu\nu}F^{\mu\nu},
\end{equation}
where
\begin{equation}
\label{12}
\varphi=\frac{\varphi_1+i\varphi_2}{\sqrt{2}},\qquad D_\mu\varphi=
\partial_\mu\varphi-ieA_\mu \varphi, \qquad F_{\mu\nu}=\partial_\mu A_\nu-\partial_\nu A_\mu.
\end{equation}
The scalar field $\varphi$ acquires non-zero vacuum expectation
values (v.e.v.)~\eqref{4}. Pointing these values along the
$\varphi_1$, we will have an infinite set of massive fields
$\varphi_1$ and massless fields $\varphi_2$ which represent the
Goldstone modes, one for each vacuum in the absence of the gauge
field $A_\mu$. The Goldstone modes are transformed into the
longitudinal degrees of freedom of the gauge field due to the
Higgs mechanism. This can be easily demonstrated by the standard
change of the field variables,
\begin{equation}
\label{13}
V_\mu=A_\mu-\frac{\partial_\mu\eta(x)}{e\langle\varphi\rangle_{k}},\qquad
\varphi=\frac{\langle\varphi\rangle_{k}+\sigma(x)}{2}\exp\left({\frac{i\eta(x)}{\langle\varphi\rangle_{k}}}\right).
\end{equation}
In terms of the variables~\eqref{13}, the quadratic part of the
Lagrangian~\eqref{11} is cast into the canonical form,
\begin{equation}
\label{14}
\mathcal{L}_{\sigma,V}^{(2)}=\frac12\partial_\mu
\sigma\partial^\mu\sigma-\frac12 M^2_\sigma\sigma^2-
\frac14F_{\mu\nu}F^{\mu\nu}+\frac12M^2_VV_\mu V^\mu+\text{const},
\end{equation}
where $M^2_\sigma$ is given by~\eqref{7} and the masses of gauge
field over different vacua become
\begin{equation}
\label{15}
M^2_{V,k}=e^2 \langle\varphi\rangle_{k}^2, \qquad k=0,1,2\dots.
\end{equation}
The relations~\eqref{7} and ~\eqref{15} demonstrate that the
infinite towers of massive vector and scalar particles emerge. It
is straightforward to generalize the model to the case of the
gauge group $SU(2)$, see Appendix~B.

Let us seek for the Nielsen--Olesen vortex solutions.
The equations of motion for the Lagrangian~\eqref{11} are:
\begin{gather}
\label{16}
D^\mu D_\mu\varphi=-\mu^2\varphi
\sin\left(\frac{2\lambda}{\mu^2}\varphi\varphi^*-2\pi b\right), \\
\partial^\nu F_{\mu\nu}=
ie(\varphi^\ast\partial_\mu\varphi-\varphi\partial_\mu\varphi^*)+2e^2A_\mu\varphi\varphi^*.
\label{17}
\end{gather}
We will consider the static case, with the gauge choice $A_0=0$.
Following the paper~\cite{nielsen} we look for a cylindrically
symmetric solution, with axis along the $z$-direction. The
corresponding ansatz is~\cite{nielsen}
\begin{equation}
\label{18}
\vec{A}(\vec{r})=\frac{\vec{r}\times\vec{e}_z}{r}A(r),
\qquad \varphi(\vec{r})=\chi(r)e^{i n\theta},
\end{equation}
where $\vec{e}_z$ denotes a unit vector along the $z$-direction and
$n$ is an integer. In addition, it is assumed that $A_\theta=A(r)$
and $A_r=A_z=0$. The flux is given by $\Phi(r)=2\pi r A(r)$ so that
the magnetic field is
\begin{equation}
\label{19}
B(r)=\frac{\Phi'(r)}{2\pi r}=\frac{(r A(r))'}{r},
\end{equation}
where the prime stays for the derivative with respect to $r$.
Inserting this ansatz to the equations of motion we arrive at
\begin{gather}
\label{20}
\frac{\left(r\chi'\right)'}{r}-\left[\left(\frac{n}{r}-e A\right)^2
+\mu^2\sin\left(\frac{2\lambda}{\mu^2}\chi^2-2\pi b\right)\right]\chi=0,\\
\left(\frac{\left(rA\right)'}{r}\right)'-2\chi^2\left(A e^2-\frac{n e}{r}\right)=0.
\label{21}
\end{gather}
The exact solution of this system cannot be obtained analytically.
We are going to find a solution with asymptotic behavior on the
spacial infinity of the type $\chi\simeq \text{const}$. It is rather
clear that the constant should be proportional to the vacuum
expectation value of the field $\varphi$. Treating $\chi$ as a
constant we get from the equation~\eqref{21}, with $C$ a constant of
integration and $K_\nu$ the modified Bessel function of the second
kind,
\begin{equation}
\label{22}
A_k(r)=\frac{n}{er}+\frac{C}{e}K_1\left(e\langle\varphi\rangle_{k}r\right)=\frac{n}{er}+
\frac{C}{e}\sqrt{\frac{\pi}{2e\langle\varphi\rangle_{k}r}}\,e^{-e\langle\varphi\rangle_{k}r}+\dots.
\end{equation}
Then the magnetic field is
\begin{equation}
\label{23}
B_k=-C\langle\varphi\rangle_{k}rK_0\left(e\langle\varphi\rangle_{k}r\right)=-
C\sqrt{\frac{\pi\langle\varphi\rangle_{k}}{2er}}\,e^{-e\langle\varphi\rangle_{k}r}+\dots.
\end{equation}
Here the dots mean the lower order terms at $r\rightarrow\infty$.
Substituting the asymptotics~\eqref{22} for $A_k(r)$ into the
equation~\eqref{20} we obtain the approximate solution
$\chi_k\simeq\langle\varphi\rangle_{k}/\sqrt{2}$,
%\begin{equation}
%\label{24}
%\chi_k\simeq\frac{\langle\varphi\rangle_{k}}{\sqrt{2}},
%\end{equation}
which defines the characteristic length $\Lambda$ (meaning the
penetration length of the magnetic field in the condensed matter
physics), $\Lambda_k=(e\chi_k)^{-1}$.
%\begin{equation}
%\label{25}
%\Lambda_k=\frac{1}{e\chi_k}=\sqrt{\frac{\lambda}{\pi e^2\mu^2(k+b)}}, \qquad k=0,1,2\dots.
%\end{equation}
In contrast to the Nielsen--Olesen theory~\cite{nielsen}, here one
formally has an infinite number of characteristic lengths, however,
the real physical penetration length is determined by the largest
one corresponding to $k=0$.

We thus see that each vacuum possesses not only its own spectrum
but also its own vortex (or monopole in the $SU(2)$ case, see
Appendix~B). As was demonstrated in~\cite{nielsen} the
vortex-lines have a flux-tube structure and move according to the
equation of motion of the Nambu dual string. A simple estimation
shows~\cite{nielsen} that the energy density of such a flux-tube
is proportional to $\langle\varphi\rangle^2$, {\it i.e.} the
energy density of the vortex-line in the $k$-th vacuum is given by
the relation~\eqref{4}. On the other hand, the dimensional
analysis of the energy functional corresponding to the
Lagrangian~\eqref{11} shows that the vortex mass is fixed,
$M_{\text{vor}}\sim\frac{\mu}{\lambda}$~\cite{nielsen}. This means
that the cross-section of the $(k+1)$-th vortex is squeezed by the
factor $k$.

\section{Quantum correction}

The interaction in the Lagrangian~\eqref{2} is non-renormalizable
in four dimensions. It is well known that for the effective field
theories the renormalizability is a desirable but not necessary
property because these theories are often considered as
approximations valid in the tree level only. Nevertheless, the
analysis of the first quantum corrections in the
non-renormalizable effective models may occasionally lead to
unexpected insights into some observable effects. In any case, it
is interesting to compare the impact of the first quantum
correction to the model~\eqref{2} with that of the
$\lambda\varphi^4$ scalar theory with the spontaneous symmetry
breaking.

A convenient formalism for the calculation of quantum corrections
is based on the effective potential. For the theory of one scalar
field, the expansion of the potential in the Planck constant is
given by ~\cite{ramond}
\begin{equation}
\label{v1}
V_{\text{eff}}=V(\varphi_{\text{cl}})+\hbar\frac{(V''(\varphi_{\text{cl}}))^2}{64
\pi^2}\left(-\frac32+\ln\frac{V''(\varphi_{\text{cl}})}{\mathcal{M}^2}\right)+\mathcal{O}(\hbar^2),
\end{equation}
where $V$ is the classical potential expanded on the field
configuration $\varphi_{\text{cl}}$ minimizing the classical
action. For the Lagrangian~\eqref{2} one has,
\begin{equation}
\label{v2}
V(\varphi_{\text{cl}})=-\frac{\mu^4}{2\lambda}\left(\cos\left(\frac{\lambda}{\mu^2}\varphi_{\text{cl}}^2-
2\pi b\right)-\cos (2\pi b)\right),
\end{equation}
where the second derivative of $V$ reads
\begin{equation}
\label{v3}
V''(\varphi_{\text{cl}})=\mu^2\left(2\frac{\lambda}{\mu^2}\varphi_{\text{cl}}^2\cos\left(2\pi
b-\frac{\lambda}{\mu^2}\varphi_{\text{cl}}^2\right)-\sin\left(2\pi
b-\frac{\lambda}{\mu^2}\varphi_{\text{cl}}^2\right)\right).
\end{equation}
In our natural units, we must of course set $\hbar=1$. The quantum
correction induces the dependence of effective
potential~\eqref{v1} on the energy scale $\mathcal{M}$. This scale
will be regarded as a new mass parameter of the model.

\begin{figure}[ht]
%\vspace{-0.8cm}
\center{\includegraphics[width=1\linewidth]{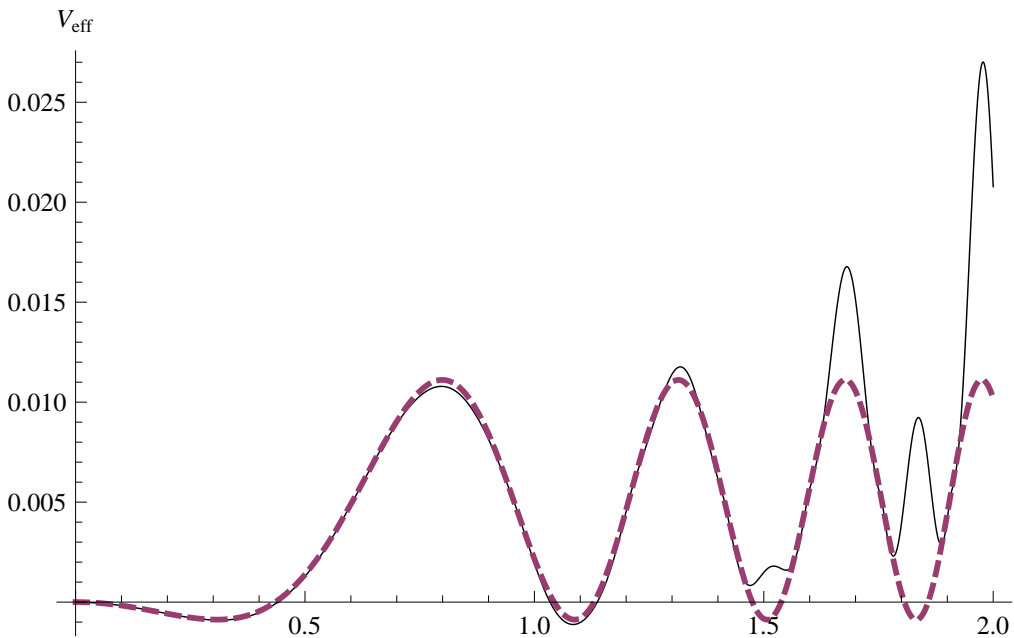}}
%\vspace{0cm}
\caption{The effective potential~\eqref{v1} (solid line)
and the classical potential~\eqref{v2} (dashed line) at $\lambda=-0.4$.
The parameters are taken from the trajectory of $f_0$-mesons in
Fig.~\ref{Fig.1}: $\mu^2=0.07$~GeV$^2$,
$b=1.41$ and we set $\mathcal{M}=0.5$~GeV.}
\label{lamb}
\end{figure}

A typical behavior of the effective potential for a particular
choice of $\lambda$ and $\mathcal{M}$ is displayed in
Fig.~\ref{lamb}. The quantum correction to the first minimum
(hence, to the mass of the ground state) is very close to the case
of the $\lambda\varphi^4$ scalar theory with the spontaneous
symmetry breaking~\cite{ramond}. But after the second minimum a
qualitative change of the picture occurs: The quantum correction
splits each minimum of the classical potential into two new nearly
degenerate minima. This means the quantum doubling of the radially
excited states. It is important that the new minima lie higher
than the first minimum. This property also has a direct
interpretation: The more massive is the radial excitation, the
more is it unstable. In order to achieve this correct physical
picture, the parameter $\mathcal{M}$ must not exceed the typical
hadronic scale, $\mathcal{M}\lesssim 1$~GeV (a concrete boundary
value slightly depends on $\lambda$).

The calculation of the first quantum correction to the vector
model of Sect.~3 is much more complicated. But our tentative
estimations show that a similar phenomenon of quantum doubling
will take place. The predicted phenomenon is difficult to observe
because of overlapping decay widths of highly excited resonances.
But it is interesting to notice that in some cases, when the
available data is rich enough and the neighboring resonances are
relatively narrow, a similar doubling is indeed observed in the
spectroscopy of the light mesons. The case of $f_2$-mesons in the
figure of Appendix~A demonstrates such a phenomenon. In terms of
the quark model, the doubling can have a simple
explanation~\cite{cluster}: Two nearly degenerate identical
resonances may represent the states with different angular
momentum $L$ of the quark-antiquark pair. For instance, the vector
mesons can have $L=0$ and $L=2$ and these states are nearly
degenerate due to the mass dependence $M_{n,L}\sim
n+L$~\cite{cluster}. The degeneracy here takes place for the
radial number $n\geq2$, {\it i.e.} beginning from the third
minimum as in Fig.~\ref{lamb}. If the so-called polarization data
is available (see Bugg's review in Ref.~\cite{phen}), even the
broad resonances can be, in principle, distinguished. An example
is given by the heaviest $\omega_3$-meson in the figure of
Appendix~A. In the case of the scalar mesons, the source of
doubling can arise from the mixing with the $s\bar{s}$ scalar
states ($s$ denotes the strange quark)~\cite{phen}. The scalar
mesons with the dominant $s\bar{s}$-component are not displayed in
Appendix~A.

\section{Discussions}

We have built an example of effective field model describing the
infinite radial spectrum of the scalar and vector particles. This
infinite spectrum can be accompanied by an infinite number of
topological vertices reminiscent of the dual strings. The particle
spectrum and topological solutions coexist and it is not clear
whether there is a deep sense in this coexistence or the classical
solutions represent just an artefact of the model. The original
Nielsen--Olesen's idea~\cite{nielsen} was to identify the
vortex-line solution with the  Nambu dual string. The quantization
of the latter had to give rise to the excited hadron spectrum. We
deal with a quite different situation: The spectrum of excitations
emerges on the classical level and each state can be accompanied
by a vortex. This approach avoids all notorious problems with
quantization of the hadron strings. The existence of vortices
might be speculatively interpreted as a footprint of real QCD
string. A precise meaning of this interpretation is then needed.
The mass of the excited state cannot be related with the mass of
the corresponding vortex since the latter mass is universal for
all vacua. The mass of the $n$-th excitation is rather
proportional to the sum of masses of $n+1$ vortices which can be
created in the vacua below the given energy scale (see the
discussions at the end of Sect.~3). In this sense, the $n$-th
radial excitation might be interpreted as "consisting of" $n+1$
vortices. This would give a "dual" explanation for the fact that,
contrary to the expectations from the QCD string models, the size
of the radial excitations do not grow with $n$. In terms of the
hadron strings, one observes a "quantization" of the energy
density in the gluon flux-tube rather than a "quantization" of its
length with a constant energy density. In other words, the
vortices could provide a classical viewpoint on the spatial
structure of the excited mesons.

From the phenomenological point of view, the model under
consideration predicts a linear Regge-like spectrum for radially
excited scalar and vector states. In the general situation, the
slopes are different in contrast to the hadron string models where
the universal slope is determined by the gluodynamics. The existing
experimental data on the light non-strange mesons seems to agree
with the approximate universality of slopes of radial meson
trajectories within the experimental errors~\cite{phen}. We can
easily incorporate this universality by setting $e^2=2\lambda$ in
the mass relations~\eqref{7} and~\eqref{15}. But this entails a
complete degeneracy of the scalar and vector states. If we associate
the scalar field $\varphi$ with the scalar isoscalar particles
($f_0$-mesons) and the vector field with the vector isoscalar states
($\omega$-mesons) this prediction is not far from the reality, see
Fig.~1.
\begin{figure}[ht]
\vspace{-0.8cm}
\center{\includegraphics[width=1\linewidth]{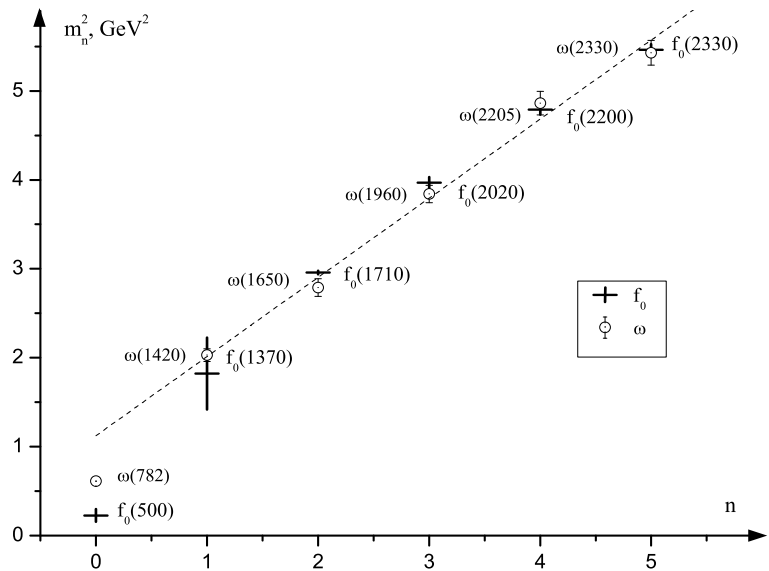}}
\vspace{-1.7cm}
\caption{The spectrum of $f_0$ and $\omega$  mesons~\cite{pdg}.
The vertical size displays the experimental
uncertainty. Other states $f_0$ reported by the Particle
Data~\cite{pdg} are assumed to belong to the $s\bar{s}$-trajectory
where the strange component dominates~\cite{phen}. The state
$\omega(2290)$~\cite{pdg} is identified with the
$\omega(2330)$-meson.}
\label{Fig.1}
\end{figure}

If we wish to keep the
universal slope but remove the degeneracy we can simply introduce
the mass term for the vector field in the Lagrangian~\eqref{11} and
consider a real scalar field. This is tantamount to adding the
vector part,
\begin{equation}
\label{26}
\mathcal{L}_v=-\frac{1}{4}F_{\mu\nu}F^{\mu\nu}+\frac12(m_v^2+e^2\varphi^2)A_\mu A^\mu,
\end{equation}
to the Lagrangian~\eqref{2}. The vector field $A_\mu$ in~\eqref{26}
can be also $SU(2)$ massive gauge field carrying the isospin one.

The fact that the field $\varphi$ is scalar does not imply that it
should describe the scalar quark-antiquark states. It looks more
natural to interpret the field $\varphi$ and its periodic
potential as an effective model for the non-perturbative gluon
vacuum in QCD. The spectrum of the Lagrangian~\eqref{2} represents
the scalar glueballs in this interpretation. The
relation~\eqref{7} gives the Regge-like form for this
spectrum\footnote{The lattice simulations seems to agree with this
prediction. In the work~\cite{meyer}, four scalar glueballs were
reported. Their spectrum is very well fitted by the
relation~\eqref{1} with the parameters: $a\approx4.5$~GeV$^2$,
$b\approx0.56$. The other lattice simulations yield less glueball
states in the scalar channel (a compilation is given
in~\cite{gregory}) but result in the same qualitative conclusion
on the scalar glueball trajectory.}. All quarkonia in the sector
of light quarks get masses mainly through interaction with the
vacuum field $\varphi$. The quark-antiquark scalar mesons should
be then treated in the same fashion as the vector ones in the
model~\eqref{26}, {\it i.e.} the simplest Lagrangian for a scalar
quarkonium $f$ is
\begin{equation}
\label{27}
\mathcal{L}_s=\frac{1}{2}\partial_\mu f\partial^\mu\! f-
\frac12(m_s^2+g_s^2\varphi^2)f^2,
\end{equation}
which should be added to the Lagrangian~\eqref{2}. Here $f$ is a
scalar or pseudoscalar field which can be both isosinglet and
isotriplet. The bare mass $m_s$ must depend on quantum numbers in
order to provide different intercepts for the radial meson
trajectories~\cite{phen}. The approximate universality of slopes
entails universality of couplings $g_s^2\simeq e^2\simeq2\lambda$.

The higher spin mesons can be also introduced by the
Lagrangian~\eqref{27} in which $f$ will denote a higher-spin (HS)
field. It represents a totally symmetric traceless tensor of rank
$J$, where $J$ denotes the spin of particle~\cite{HS}. The kinetic
and mass term will have additional contributions due to different
ways for contracting the Lorentz indices. The hadron strings predict
the Regge behavior $M^2\sim J$ which is well seen in the observed
spectrum of hadrons~\cite{phen,klempt}. This behavior can be
incorporated by assuming that the mass term is proportional to the
number of physical degrees of freedom $2J+1$ ({\it i.e.} each degree
of freedom contributes to the bare mass as the scalar field
in~\eqref{27}). The model will contain interactions of HS fields
with the excitations of the vacuum field. This is a problem because
a self-consistent theory of interacting HS fields is absent.
However, if we are interested in the mass spectrum which is supposedly
determined by the quadratic part of the action, this problem does
not appear.

The models discussed above should be viewed as effective models for
QCD in the large-$N_c$ limit (because the number of resonances is
infinite and the zero-width approximation is implied). They are
complementary to the usual effective field theories for the strong
interactions (the sigma-models, chiral perturbation theory, {\it
etc}). The precise sense of complementarity is as follows. The
standard effective models are low-energy ones, {\it i.e.} they are
usually defined below the characteristic scale of the spontaneous
chiral symmetry breaking (CSB) --- about 1~GeV --- and their input
parameters are supposed to arise from integration of the degrees of
freedom above this scale. In our case, the situation is opposite ---
by assumption, the input parameters (first of all the bare masses
in~\eqref{26} and~\eqref{27}) originate from integration of the
low-energy degrees of freedom below 1~GeV. A phenomenological
support for such a viewpoint comes from the observation that the
linear ansatz~\eqref{1} for the radial meson spectrum is a
reasonable approximation for the excited states, while the ground
ones below 1~GeV deviate substantially from the linear trajectory.
This is seen, for example, in Fig.~1. The deviation is the most
pronounced in the pseudoscalar channel~\cite{phen}. Below 1~GeV, the
physics of strong interactions is shaped by the CSB. The building of
the low-energy effective theories is therefore guided by the chiral
symmetry. Above 1~GeV, the CSB is not crucial~\cite{glozman} and the
physics is governed by the gluodynamics. QCD in the large-$N_c$
limit represents (assuming that confinement persists in this limit)
a theory of an infinite number of narrow and stable non-interacting
mesons and glueballs~\cite{hoof} which should appear in the
classical effective action. The models considered above satisfy
these basic properties.

The interest to the physics of radially excited hadrons has
recently been raised due to a fast development of the
five-dimensional holographic approach to QCD. Specifically, the
Regge-like spectrum~\eqref{1} is reproduced within the so-called
Soft-Wall model~\cite{son2}. Exploring the gauge/gravity duality
from the string theory, the holographic models identify the
excited hadrons with the Kaluza-Klein excitations. Our approach is
alternative to the holographic one and also attempts to realize a
certain concept of duality. Namely, the proposed models are
constructed in the spirit of the so-called "dual QCD" approach
born by the Nielsen--Olesen paper~\cite{nielsen} (the history and
achievements of this line of research are briefly reviewed in
Ref.~\cite{baker}). Within this approach, one tries to reformulate
QCD in the long-distance limit as some weakly-coupled "dual"
theory which should describe the non-perturbative physics of the
strong interactions. Unfortunately, no connection between the
"dual" variables and the QCD ones has been derived. Thus any model
based on the "dual QCD" represents a purely phenomenological
approach. Being aware of theoretical difficulties we would dare
nevertheless to outline a possible way for finding certain
connections with QCD. It should be reminded that the concept of
duality (in the sense of equivalence of two theories plus the
strong--weak coupling correspondence) first emerged in the
two-dimensional field theories. A spectacular example is given by
the duality between the quantized sine-Gordon model and the
Thirring model discovered by Coleman~\cite{ra}. It would be
extremely interesting to establish an approximate fermionic
theory, bosonization of which leads to our models. This may shed
light on the problem of emergence of considered models from QCD
since such a fermionic theory can serve as a toy-model for the
effective theory of QCD obtained after integrating out the gluonic
degrees of freedom.

\section{Conclusions and outlook}

We have proposed a novel Lagrangian description for the radially
excited mesons. The main assumption of our scheme is that the
non-perturbative vacuum of QCD has periodic structure in energy
scale. The radially excited states appear as elementary
excitations over different non-equivalent vacua. The Regge-like
form of their spectrum represents, by construction, the main
feature of the model and is a consequence of the vacuum
periodicity. Each vacuum contains a non-trivial topological
configuration which accompanies the corresponding radial
excitation. To some extend, the considered models lead to the
quantum doubling of the radially excited states and to the
linearity of the scalar glueball trajectory. Both phenomena have a
certain phenomenological support.

There are many questions which can be (and hopefully will be)
addressed within the introduced field models. We mention some of
them. (i) The precise relation between the excited mesons and the
accompanying classical solutions. (ii) The construction of
equivalent fermionic models. (iii) The calculation of the first
quantum correction to the vector spectrum. (iv) The influence of
non-trivial external conditions (especially the non-zero
temperature and finite density) on the whole set of the radially
excited mesons. (v) Since the sine-Gordon model has interesting
applications in the condensed matter physics and the
Nielsen--Olesen theory does in cosmology (cosmic strings), the
model of Sect.~2 may have applications in the first field and the
model of Sect.~3 may have in the second one.

\section*{Acknowledgments}

The work was partially supported by the Saint Petersburg State
University grants ¹ 11.38.660.2013 and 11.48.1447.2012, by the
RFBR grants 13-02-00127-a and 12-02-01121-a, and by the Dynasty
Foundation.

\section*{Appendix A: Meson clusters}

In this Appendix, we reproduce the figure from
review~\cite{clust_rev} showing the known spectrum of the light
non-strange mesons~\cite{pdg}. The experimental uncertainties are indicated
and purely established states are not shaded. The positions of
meson clusters (first described in Ref.~\cite{Cluster}) are marked
by the vertical dashed lines.

\begin{figure*}
\vspace{-4.5cm}
\hspace{-3cm}
\resizebox{1.5\textwidth}{!}{\includegraphics{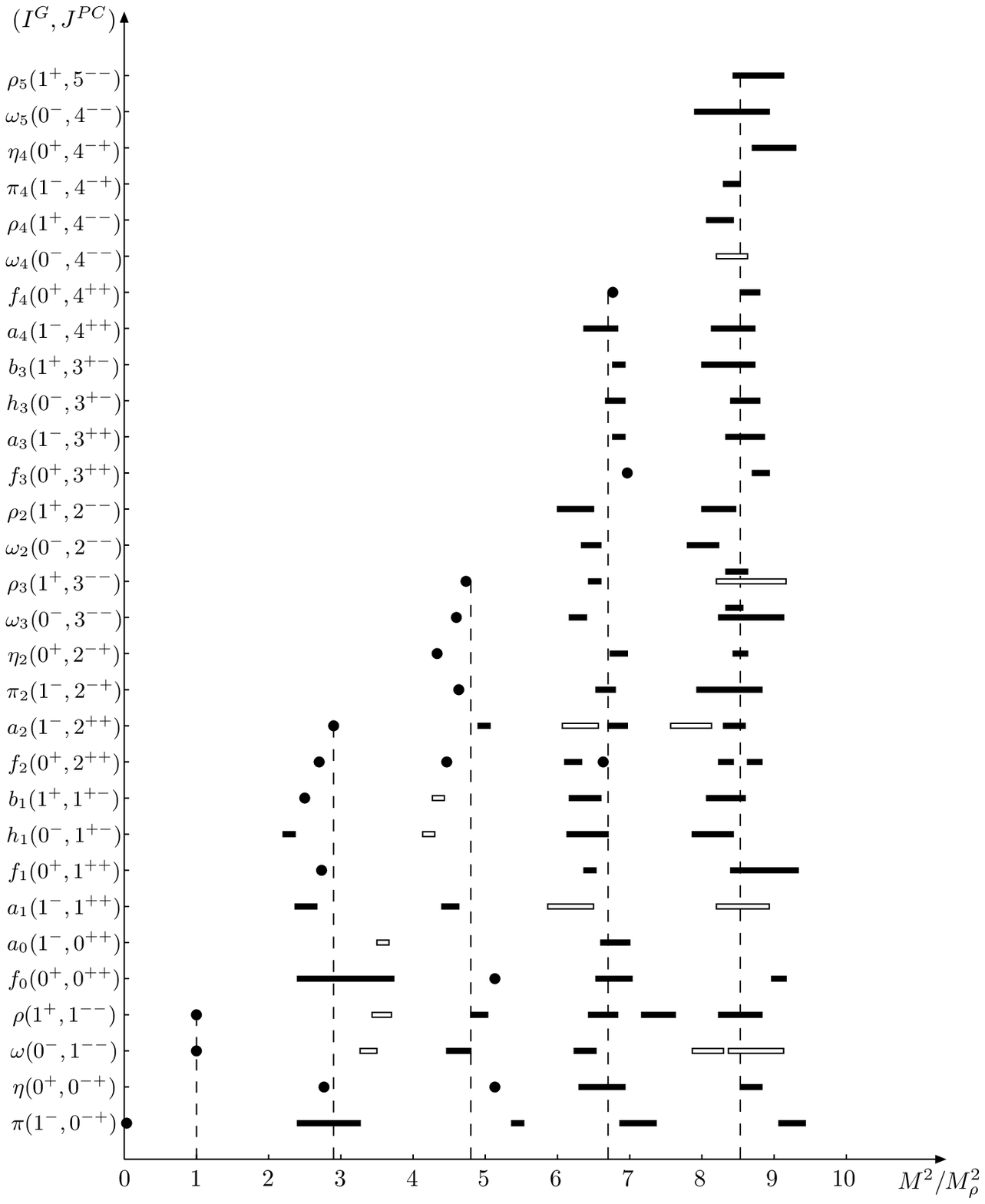}}
\end{figure*}

\newpage
\section*{Appendix B: Non-Abelian extensions}

The construction of the non-Abelian extensions for the model of
Sect.~3 is straightforward. We will consider the case of the gauge
group $SU(2)$. Let the scalar field $\varphi$ transform according
to the real vector representation. The extension of the
Lagrangian~\eqref{11} takes then the form
\begin{multline}
\mathcal{L}=-\frac{1}{4}F_{\mu\nu}^a
(F^{\mu\nu})^a+\frac12(D^\mu \varphi^a)(D_\mu
\varphi^a)\\
+\frac{\mu^4}{2\lambda}\left[\cos\left(\frac{\lambda}{\mu^2}\varphi^a\varphi^a-2\pi
b\right)-\cos(2\pi b)\right],
\label{vln}
\end{multline}
where $a=1,2,3$ and the field strength and the covariant
derivative are defined by
\begin{gather*}
F_{\mu\nu}^a=\partial_\mu A_\nu^a-\partial_\nu A_\mu^a+g\varepsilon^{abc} A_\mu^b A_\nu^c,\\
D_\mu \varphi^a=\partial_\mu \varphi^a+g\varepsilon^{abc}A_\mu^b\varphi^c.
\end{gather*}

The scalar field acquires the non-zero v.e.v.
$$
\varphi^a\varphi^a=\langle\varphi\rangle_{k}^2.
$$
Consider the small fluctuation of the field $\varphi_a$ near its
v.e.v.
$$
\varphi_a=(\varphi_0)_a+\sigma_a.
$$
The gauge freedom
allows to point the vector $\varphi_0$ along the third axis,
\begin{equation}
(\varphi_0)_a=\langle\varphi\rangle_{k} \delta_{a3}.
\label{vvac}
\end{equation}
Substituting~\eqref{vvac} into the Lagrangian~\eqref{vln} and
keeping only the terms quadratic in fields, we arrive at
\begin{multline}
\mathcal{L}_{\sigma,A}^{(2)}=\frac12\partial_\mu
\sigma\partial^\mu\sigma-\frac12
M^2_\sigma\sigma^2-\frac{1}{4}{\mathcal F_{\mu\nu}^a}({\mathcal
F^{\mu\nu}})^a\\
+\frac{1}{2}M^2_{A}\left((A_\mu^1)^2+(A_\mu^2)^2\right)+\text{const},
\label{vln1}
\end{multline}
where
\begin{gather*}
\mathcal F_{\mu\nu}^a=\partial_\mu A_\nu^a-\partial_\nu A_\mu^a,\\
M^2_{A}=M^2_{A^1}=M^2_{A^2}=g^2\langle\varphi\rangle_{k}^2,
\end{gather*}
and the gauge boson $A_\mu^3$ (corresponding to the unbroken symmetry
with respect to rotation around the third axis) remains massless.

Another realization of the Higgs model with the $SU(2)$ gauge
group is given by the complex spinor representation of the scalar
field,
$$
\varphi=\left(\begin{array}{c}
              \frac{\varphi_1+i\varphi_2}{\sqrt{2}} \\
              \frac{\varphi_3+i\varphi_4}{\sqrt{2}} \\
            \end{array}\right).
$$
The corresponding Lagrangian is
\begin{multline}
\mathcal{L}=-\frac{1}{4}F_{\mu\nu}^a (F^{\mu\nu})^a+(D^\mu
\varphi)^\dag(D_\mu
\varphi)\\
+\frac{\mu^4}{2\lambda}\left[\cos\left(\frac{2\lambda}{\mu^2}\varphi\varphi^\dag
-2\pi b\right)-\cos(2\pi b)\right],
\label{isosp}
\end{multline}
where
\begin{gather*}
F_{\mu\nu}^a=\partial_\mu A_\nu^a-\partial_\nu A_\mu^a+g\varepsilon^{abc} A_\mu^b A_\nu^c,\\
D_\mu \varphi=\partial_\mu \varphi-ig\frac{\tau^a}{2}A_\mu^a\varphi.
\end{gather*}
The classical configurations delivering minimum to the
effective potential satisfy the condition
$$
\varphi\varphi^\dag=\frac{\langle\varphi\rangle_{k}^2}{2}.
$$
The v.e.v.'s of the fields are $A_\mu^a=0$ and
$$
\varphi=\left(\begin{array}{c}
              0 \\
              \frac{\langle\varphi\rangle_{k}}{\sqrt{2}} \\
            \end{array}\right).
$$
Consider the small scalar excitations over the vacua (the unitary
gauge is chosen),
$$
\varphi=\left(\begin{array}{c}
              0 \\
              \frac{\langle\varphi\rangle_{k}+\eta}{\sqrt{2}} \\
            \end{array}\right).
$$
The quadratic in fields part of the Lagrangian takes the form
\begin{equation}
\mathcal{L}_{\sigma,A}^{(2)}=\frac12\partial_\mu
\sigma\partial^\mu\sigma-\frac12
M^2_\sigma\sigma^2-\frac{1}{4}{\mathcal F_{\mu\nu}^a}({\mathcal
F^{\mu\nu}})^a+\frac12M^2_{A^a}A_\mu^a(A^\mu)^a+\text{const},
\end{equation}
where the spectrum for $M^2_\sigma$ coincides with~\eqref{7} and
the vector spectrum is
\begin{equation}
M^2_{A^a}=\frac{g^2 \langle\varphi\rangle_{k}^2}{4}, \qquad
k=0,1,2...,\qquad a=1,2,3.
\end{equation}

The analysis of classical solution in the isovector case results
in the 't~Hooft--Polyakov monopoles. We will reproduce briefly the
standard derivation of those solutions.

Consider the static case and set $A_0^a(\vec{x})=0$ for any
$\vec{x}$ and $a$. The equations of motion following from the
Lagrangian~\eqref{vln} are
\begin{gather}
\label{M3}
D^i D_i\varphi^a=\mu^2\varphi^a
\sin\left(\frac{\lambda}{\mu^2}\varphi^b\varphi^b-2\pi b\right), \\
D_i(F^{i j})^a=
-g\varepsilon^{abc}(D^j\varphi^b)\varphi^c.
\label{M4}
\end{gather}
Let us look for the solutions in the form of the hedgehog
ansatz~\cite{polyakov},
\begin{equation}
\varphi_a=x_a \frac{u(r)}{r}, \qquad
A_i^a=a(r)\varepsilon_{iab}x_b,
\end{equation}
which after the substitution to Eqs.~\eqref{M3} and~\eqref{M4} leads
to the equations,
\begin{gather}
a''+4r^{-1}a'-3ga^2-g^2r^2a^3-g^2u^2a=gu^2r^{-2}, \\
u''\!+\!2r^{-1}(u'\!-\!r^{-1}u)\!-\!2gau\!-\!2g^2a^2r^2u\!-\!\mu^2u\sin\left(\!\frac{\lambda}{\mu^2}u^2\!-\!2\pi b\!\right)\!=\!0.
\end{gather}
These equations cannot be solved analytically but it is easy to
find the asymptotics of the solutions at large distances,
\begin{equation}
r\rightarrow\infty:\qquad u(r)\propto \langle\varphi\rangle_{k}, \qquad
a(r)\propto -\frac{1}{gr^2}.
\end{equation}

In order to extract the physical fields 't~Hooft introduced the
following gauge-invariant definition for the field strength
tensor~\cite{Hooft_mon},
\begin{equation}
\mathcal{F_{\mu\nu}}=\frac{1}{|\varphi|} \varphi_a (F^{\mu\nu})^a
-\frac{1}{g|\varphi|^3}\varepsilon^{abc}\varphi_a(D_\mu\varphi_b)(D_\nu\varphi_c).
\end{equation}
If one has $\varphi_a=|\varphi|(0,0,1)$ in some domain for any
gauge transformation then $\mathcal{F_{\mu\nu}}=\partial_\mu
A_\nu^3-\partial_\nu A_\mu^3$. The asymptotics of
$\mathcal{F_{\mu\nu}}$ in the spatial infinity is
$$
\mathcal{F_{\mu\nu}}\rightarrow- \frac{1}{g r^3}\varepsilon_{\mu\nu a}x_a.
$$
Here the symbol $\varepsilon$ is zero if any of four-dimensional
indices takes the value 4. We thus obtain the expression for the
magnetic field corresponding to the magnetic charge
$g_m=1/g$,
\begin{equation}
B_a(\vec{x})=\frac{x_a}{g r^3}, \qquad r\rightarrow\infty.
\end{equation}
It is seen that the vector field extends to the spatial infinity.
This is related with the existence of the massless mode
(see~\eqref{vln1}) which can penetrate via the Bose-condensate of
the scalar field. The given observation allows to prove (see, {\it
e.g.}, Ref.~\cite{polyakov}) that the hedgehog solution represents
the Dirac monopole, {\it i.e.} the vector-potential has the form
$$
A_r=A_\theta=0, \qquad A_\varphi=\frac{g_m}{r}\frac{1-\cos \theta}{\sin \theta}.
$$

As in Sect.~3, we thus conclude that for each vacuum there exists
its own non-trivial classical field configuration --- the
't~Hooft--Polyakov monopole in the case under consideration. The
magnetic field created by these monopoles in the spatial infinity
is, however, equal for any vacuum.

In the case of the isospinor Higgs field --- the
model~\eqref{isosp} --- all gauge fields are massive and located
in a domain of the size approximately $1/M_A$. The "magnetic"
field in the spatial infinity does not appear in this model and
the isospinor hedgehog is not a magnetic monopole. In addition,
the isospinor hedgehog is unstable~\cite{polyakov}.

\end{document}